\documentclass[10pt,preprintnumbers,aps,amssymb,nofootinbib,amsmath,superscriptaddress,twocolumn,prl]{revtex4-1}
\usepackage{epsfig,epsf}
\usepackage{bm} 
\newcommand{\beq}{\begin{equation}}
\newcommand{\beql}[1]{\begin{equation}\label{#1}}
\newcommand{\eeq}{\end{equation}}
\def\bal#1\gal{\begin{align}#1\end{align}}
\newcommand{\ball}[1]{\bal\label{#1}}
\newcommand{\eq}[1]{(\ref{#1})}
\newcommand{\fig}[1]{Fig.~\ref{#1}}

\newcounter{topiccounter}
\renewcommand{\b}[1]{{\bm #1}} 

\newcommand{\as}{\alpha_s}

\newcommand{\im}{\,\mathrm{Im}\,}
\newcommand{\real}{\,\mathrm{Re}\,}

\newcommand{\aver}[1]{\left\langle #1 \right\rangle}

\begin{document}

\title{Violation of  geometric scaling in DIS due to Coulomb corrections}

\author{Kirill Tuchin}

\affiliation{Department of Physics and Astronomy, Iowa State University, Ames, IA, 50011}

\date{\today}


\begin{abstract}
We compute the Coulomb correction to the total and diffractive cross sections for virtual photon scattering off a heavy nucleus at low $x$.   We show that it violates  the geometric scaling in a wide range of photon virtualities and is weakly $x$-independent. In heavy nuclei at low $Q^2$ the Coulomb correction to the total and diffractive cross sections is about 20\% and 40\% correspondingly. 
\end{abstract}

\maketitle


\section{Introduction}\label{sec:Int}

A pivotal property of the low $x$ semi-inclusive DIS on proton and nuclear targets is geometric scaling  of the total $\gamma^*p$ and $\gamma^*A$ cross sections \cite{Stasto:2000er}, which means scaling with a dimensionless ratio $Q^2/Q_s^2(x)$, where $Q^2$ is photon virtuality, $x$ is Bjorken variable and $Q_s^2(x)$ is the saturation momentum. Geometric scaling --  a fundamental property of high energy QCD  \cite{Gribov:1984tu} -- is a most clear manifestation of  a highly coherent color field, which has a typical transverse momentum  scale $Q_s(x)$. In \cite{Levin:1999mw,Levin:2000mv,Levin:2001cv,Iancu:2002tr}  it was derived from the  low $x$ evolution equation of QCD \cite{Balitsky:1995ub,Kovchegov:1999yj}. 
The coherent color field is made up mostly of gluons, which cannot directly couple to the virtual photon. Therefore, the leading DIS channel at low $x$ is a fluctuation of the virtual photon into a $q\bar q$ pair, which is a color and electric dipole, that subsequently interacts with the color field of the target. 

Predictions of the perturbation theory are most robust for DIS off a heavy nucleus $A\gg 1$ because  $\as^2 A^{1/3}\sim 1$ serves as a convenient resumation parameter. Additionally, the color field strength is boosted by a large factor $A^{1/3}$. Thus, DIS off a heavy nucleus is considered to be the best tool to probe the low $x$ nuclear structure and dynamics. Experimental facilities capable of performing such experiments, for example the Electron Ion Collider, are being actively developed.  

A large-$A$ nucleus also carries strong electric charge $eZ$. Elastic scattering amplitude of the $q\bar q$ dipole off the nuclear Coulomb field is proportional  $\alpha Z$, which is of order one for a heavy nucleus. Therefore, the cross section for DIS off a heavy nucleus also receives a substantial contribution from electromagnetic interactions of the $q\bar q$ dipole with the nucleus, which is known as the Coulomb correction.  Since the typical scale of the nuclear electromagnetic field is obviously different from the saturation momentum, the Coulomb correction violates the geometric scaling. We will argue that this correction is large at low $x$ and small $Q^2$, which is precisely the region that will be probed by the EIC and similar experiments. The main goal of this letter is to demonstrate the importance of the Coulomb correction in DIS off heavy nuclei and to investigate it as a function of $Q^2$, $x$ and $A$. Non-negligible Coulomb corrections at medium $x$ were recently discussed in \cite{Solvignon:2009it}.

\section{Total cross section}\label{sec:a}

At low $x$ the total $\gamma^*A$ cross section can be expressed in terms of the total dipole--nucleus cross section $\hat \sigma$ as follows (see e.g.\ \cite{Kovchegov:2012mbw})
\ball{a11}
\sigma_{T/L}(x,Q^2)= \frac{1}{4\pi}\int_0^1 dz \int d^2r\, \Phi_{T/L}(r,z)\,\hat \sigma(x,r)\,,
\gal  
where  $Q^2$ is the photon virtuality.  The light-cone wave functions for transverse and longitudinal polarizations of photon are given by 
\ball{a13}
\Phi_{T}= \sum_f\frac{2\alpha N_c}{\pi}\left\{ [z^2+(1-z)^2]a^2 K_1^2(ar)\right. &\nonumber \\
\left. +m_f^2 K_0^2(ar)\right\}\,, &\\
\Phi_L= \sum_f\frac{2\alpha N_c}{\pi} \, 4Q^2z^2(1-z)^2K_0^2(ar)\,, &\label{a14}
\gal
where  $m_f$ is quark mass, $z$ is the fraction of the photon's light-cone momentum  carried by the quark, $r$ is the size of the $q\bar q$ dipole in the transverse plane and $a^2= z(1-z)Q^2+m_f^2$. The relationship between  the cross section $\sigma=\sigma_T+\sigma_L$ and $F_1$, $F_2$ structure functions is non-trivial due to large Coulomb corrections to the leptonic tensor  \cite{Kopeliovich:2001dz}.

In order to calculate the Coulomb correction to the total $\gamma^*A$ cross section 
we employ the Glauber-Mueller model \cite{Glauber:1987bb,dip,Bjorken:1970ah} which takes into account multiple scatterings of the $q\bar q$ dipole in the nucleus. Let $\Gamma_\text{s}$ and $\Gamma_\text{em}$ be QCD and QED contributions to the dipole--nucleon elastic scattering amplitude. Average over the nucleus wave function can be calculated using the thickness function $T(\b b)$ as follows
\bal
\aver{\Gamma_\text{s/em}(\b b)}&= 
\frac{1}{A} \int d^2b_a\, T_A(\b b_a)\,  \Gamma_\text{s/em}(\b b-\b b_a)\,,\label{a15}
\gal 
where $\b b$ and $\b b_a$ impact parameters of  the dipole and a nucleon correspondingly.   According to the optical theorem, the dipole--nucleus cross section reads
\ball{a19}
\hat\sigma&=2\int d^2b\real \left\{ 1- \exp\left[- A \aver{i\Gamma_\text{s}} -Z\aver{i\Gamma_\text{em} }\right]\right\}\\
&= 2\int d^2b \left\{ 1- \cos[Z \aver{\real i\Gamma_\text{em}  }]\exp[- A \aver{\im i\Gamma_\text{s}}]\right\}\,,\label{a20}
\gal
where we neglect a small real part of $i\Gamma_\text{s}$ and a small imaginary part of $i\Gamma_\text{em}$. Integrals in \eq{a15} and \eq{a19} can be analytically calculated in a simple but quite  accurate ``cylindrical nucleus" model (see e.g.\ \cite{Kovchegov:2001sc,Kharzeev:2004yx}), which approximates   the nuclear thickness function by the step function, viz.\  $T(b)=2R_A$ if $b<R_A$ and  zero otherwise. The result is \cite{Tuchin:2013eya}
\ball{a21}
&\hat\sigma(x,r)=\hat\sigma_\text{s}(x,r)+ \hat\sigma_\text{em}(x,r)\,,\\
&\hat\sigma_\text{s}(x,r)= 2\pi R_A^2 \left\{ 1-\exp\left[-\frac{1}{4}\tilde Q_s^2(x) r^2\right]\right\},\label{a22}\\
&\hat\sigma_\text{em}(x,r)= 4\pi r^2 (\alpha Z)^2 \ln\frac{W^2}{4m_f^2m_NR_A},\label{a23}
\gal
where $m_N$ is nucleon mass, $W$ is the $\gamma^* A$  center-of-mass energy given by $W^2=Q^2/x+m_N^2$ and $\tilde Q_s^2$ is the \emph{quark} saturation momentum. 

Logarithm  that appears in \eq{a23} is the result of integration over the impact parameter from $R_A$ up to a cutoff $b_\text{max}$, which delimits the region of validity of the Weizs\"acker-Williams approximation. It is given by $b_\text{max}=\max\{W^2z(1-z)/(m_N(m_q^2+\b k^2))\}$, where $\b k$ is the quark's transverse momentum \cite{Tuchin:2009sg}. The largest size of the $q\bar q$ dipole, corresponding to the smallest $\b k$, is  $\sim 1/m_f$  due to the confinement. For that reason  $b_\text{max}$, and hence \eq{a23},  depends on the constituent quark mass $m_f$ rather than on the much smaller current quark mass $m_q$.

Eqs.~\eq{a21}--\eq{a23} are derived in the quasi-classical approximation where the quark saturation momentum $\tilde Q_s^2$, and hence the QCD contribution to the total cross section, is $x$-independent. At lower $x$, such that $\as \ln(1/x)\sim 1$, the QCD quantum evolution effects become important and are described by the BK equation \cite{Balitsky:1995ub,Kovchegov:1999yj}. It emerges form the solution to the BK equation that the saturation momentum acquires $x$-dependence in the form $\tilde Q_s^2\sim A^{1/3}x^{-\lambda}$, where $\lambda$ is a certain positive number \cite{Levin:2001cv}. The functional form of the dipole cross section is also evolving with $x$;  \eq{a22} in that case is the initial condition.  Several phenomenological models were suggested to describe the evolved dipole cross section. We will follow the GBW model \cite{GolecBiernat:1998js} which retains the functional form of \eq{a22} while models the saturation momentum according to \eq{e11}.
If we neglect the electromagnetic term \eq{a23} and use \eq{a22} in \eq{a11}, then we immediately observe that the total $\gamma^*A$ cross section exhibits the geometric scaling. This is because $x$-dependence arises only through the combination $r^2Q_s^2(x)$, and the dipole size $r$ is determined by $1/Q$ (for $Q^2\gg m_f^2$). 

That the Coulomb correction violates the geometric scaling is evident from \eq{a23} which, being an electromagnetic contribution,  does not depend on the strength of the color field determined by $\tilde Q_s^2$.  Unlike the QCD term \eq{a22}, the QED one \eq{a23} does not evolve much with $x$. Indeed, $\Gamma_\text{s/em}\sim (1/x)^{1+\Delta_\text{s/em}}$, where the intercept $\Delta_\text{s/em}$ incorporates the evolution effect.  In the leading-log oder in QCD $\Delta_\text{s}=4\ln 2(\as N_c/\pi)$ \cite{Balitsky:1978ic,Kuraev:1977fs}, while in QED  $\Delta_\text{em}= (11/32) \pi \alpha^2$ \cite{Gribov:1970ik,Mueller:1988ju}. Because $\Delta_\text{em}\ll \Delta_\text{s}$ we can neglect the effect of the QED evolution.

Substituting \eq{a23} into \eq{a11} and integrating over $r$ we obtain the following analytic expression for the Coulomb correction to the total $\gamma^*A$ cross section
\ball{b11}
\sigma_{\text{em},T/L}&= (\alpha Z)^2\ln\frac{W^2}{4m_f^2m_NR_A} \sum_f \frac{4\alpha N_c}{3m_f^2}\, g_{T/L}(\eta)\,,
\gal
where $\eta= Q/m_f$ and 
\ball{b13}
g_T(\eta)&=\left[ 4 \left(\eta ^4+7 \eta ^2+8\right) \tanh ^{-1}\left(\frac{\eta  \sqrt{\eta ^2+4}}{\eta ^2+2}\right)\right.    \nonumber\\
&\left.
-2 \eta  \sqrt{\eta ^2+4} \left(\eta
   ^2+8\right)\right]\left[\eta ^3 \left(\eta ^2+4\right)^{3/2}\right]^{-1},\\
g_L(\eta)&=4\left[ \eta  \sqrt{\eta ^2+4} \left(\eta ^2+6\right)   \right.\nonumber\\
&\left.
+4 \left(\eta ^2+3\right) \ln \frac{\eta -\sqrt{\eta
   ^2+4}}{\eta+\sqrt{\eta ^2+4}}\right]\nonumber\\
   &\times  \left[ \eta ^3 \left(\eta ^2+4\right)^{3/2}\right]^{-1}.\label{b15}
\gal  
Obviously, \eq{b11} with \eq{b13},\eq{b15} does not scale with $\tilde Q_s^2/Q^2$.

The QCD contribution can be  estimated  analytically only at very large and very small photon virtuality (as compared to the saturation momentum). We derive the following asymptotic   
expressions for the relative size of electromagnetic contribution  compared the total $\gamma^*A$ cross section: 
\ball{c17}
\frac{\sigma_{\text{em}}}{\sigma_{\text{s}}}&=
\frac{8\ln \frac{Q^2}{m_f^2}}{\tilde Q_s^2R_A^2\ln \frac{Q^2}{\tilde Q_s^2}}
(\alpha Z)^2\ln\frac{W^2}{4m_f^2m_NR_A} \,,
\gal
when $m_f^2\ll Q_s^2\ll Q^2$. This ratio increases logarithmically with $Q^2$, but decreases at low $x$  as $x^{\lambda}$ (modulo logarithms). We therefore expect that in this kinematic region electromagnetic interactions of the $q\bar q$ are small at very low $x$. The situation is remarkably different at semi-hard momenta where the ratio of QED and QCD contribution reads
\ball{c21}
\frac{\sigma_{\text{em}}}{\sigma_{\text{s}}}&=\frac{8\ln \frac{Q^2}{m_f^2}}{Q^2R_A^2}
(\alpha Z)^2\ln\frac{W^2}{4m_f^2m_NR_A} \,,
\gal 
when $m_f^2\ll Q^2\ll \tilde Q_s^2$.
We see that since $W^2\sim Q^2/x$, the relative size of the electromagnetic contribution slowly increases  as $\ln(1/x)$. Nuclear dependence of \eq{c17} is given by $Z^2/A$ (modulo logarithms), while that of \eq{c21} by $Z^2/A^{2/3}$, which indicates that in the saturation region \eq{c21} the relative electromagnetic contribution is enhanced by $A^{1/3}$ as compared to the hard perturbative region \eq{c17}. The number of nucleons  $A$
increases with the number of protons $Z$ in a nucleus as $A= \phi(Z) Z$, where $\phi\approx 2-3$ is a slowly increasing function of $Z$. Therefore, the above ratios are monotonically increasing functions of $Z$, indicating the enhancement of the Coulomb correction for heavy nuclei.

\begin{figure}[ht]
     \includegraphics[width=8cm]{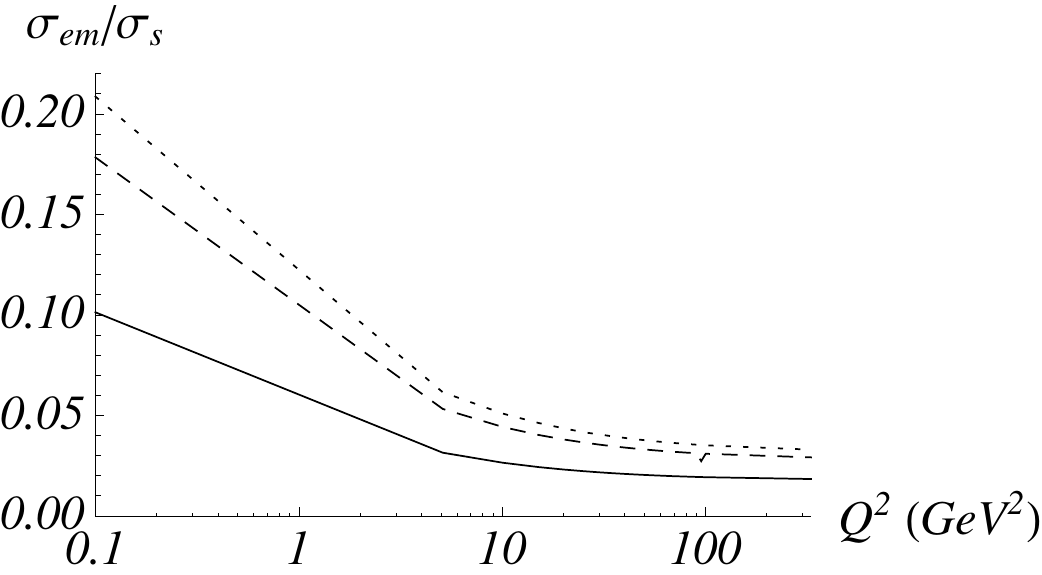}
 \caption{Ratio of QED and  QCD contributions to the total $\gamma^*A$ cross section at $x=10^{-4}$ as a function of $Q^2$ for silver (solid line), gold (dashed line) and uranium (dotted line) nuclei.  }
\label{fig1}
\end{figure}

\begin{figure}[ht]
     \includegraphics[width=8cm]{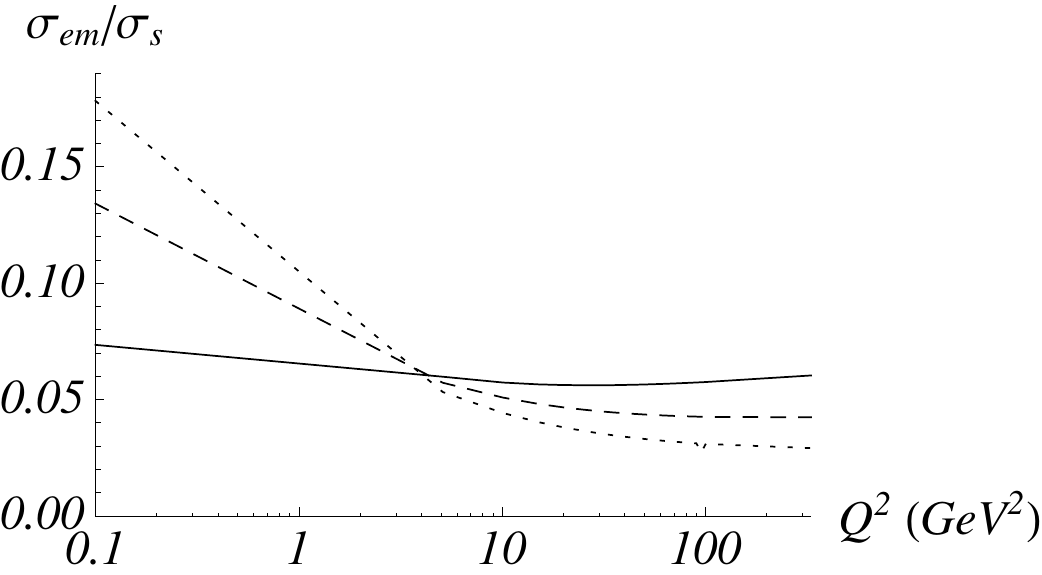} 
 \caption{Ratio of QED and  QCD contributions to the total $\gamma^*A$ cross section as a function of $Q^2$ for gold nucleus   at $x=10^{-2}$ (solid line), $x=10^{-3}$ (dashed line), $x= 10^{-4}$ (dotted line).}
\label{fig2}
\end{figure}

\section{Diffractive cross section}\label{sec:d}

Total diffractive cross section corresponds to  elastic scattering of color dipole on the nucleus. It can be written as 
\ball{d11}
\sigma^{\text{diff}}_{T/L}(x,Q^2)= \frac{1}{4\pi}\int_0^1 dz \int d^2r\, \Phi_{T/L}(r,z)\,\hat \sigma^\text{el}(x,r)\,,
\gal  
where the total elastic dipole--nucleus cross section reads
\ball{d13}
&\hat \sigma^\text{el}(x,r)=\int d^2b \left| 1- \exp\left[- A \aver{i\Gamma_\text{s}} -Z\aver{i\Gamma_\text{em} }\right]\right|^2
\gal
Following the same steps that led from \eq{a20} to \eq{a21}--\eq{a23} (details can be found in \cite{Tuchin:2013eya}), we derive
\ball{d25}
\hat \sigma^\text{el}(x,r) =\hat \sigma_\text{s}^\text{el}(x,r)+ \hat \sigma_\text{em}(x,r)\,,
\gal
where $\hat \sigma_\text{em}$ is the QED contribution given by \eq{a23}, while  the QCD contribution is
\ball{d27}
\hat \sigma_\text{s}^\text{el}(x,r)=\pi R_A^2 \left\{ 1-\exp\left[-\frac{1}{4}\tilde Q_s^2(x) r^2\right]\right\}^2\,.
\gal

Similarly to the total cross section we find  the following asymptotic relations between the QCD contributions to the total and diffractive cross sections. 
\ball{d31}
&\sigma_\text{s}=\frac{\ln\frac{Q^2}{\tilde Q_s^2} }{\ln 2}\,\sigma^\text{diff}_\text{s}\,,\quad Q^2\gg \tilde Q_s^2\,,\\
&\sigma_\text{s}=2\, \sigma^\text{diff}_\text{s}\,,\quad  \tilde Q_s^2 \gg Q^2\,.
\gal
Thus, the relative importance of the Coulomb correction in the total cross section is larger than in the diffractive one. Indeed, the QCD contribution to the diffractive cross section is obviously smaller that the total one (being part of it), while the QED contribution is the same.

\section{Numerical analysis}\label{sec:e}

To obtain a quantitative estimate of the Coulomb correction we perform a numerical calculation using \eq{a11}--\eq{a23} and \eq{d11}--\eq{d27}. The saturation momentum is parameterized according to the Golec-Biernat--Wusthoff model \cite{GolecBiernat:1998js}  in which  
\ball{e11}
\tilde Q_s^2= Q_0^2\left(\frac{x_0}{x}\right)^\lambda\,,
\gal
where  $Q_0=1$~GeV, $x_0=3.04\cdot 10^{-4}$, $\lambda=0.288$ and effective proton radius $R_p=3.1$~GeV$^{-1}$ are parameters fitted to the low $x$  DIS data. Nuclear radius is given by $R_A= R_pA^{1/3}$. We sum over three light quark flavors with constituent masses $m_f=140$~MeV. Since $W=Q^2/x$  the cross sections are functions of $x$ and $Q^2$. 

\begin{figure}[ht]
     \includegraphics[width=8cm]{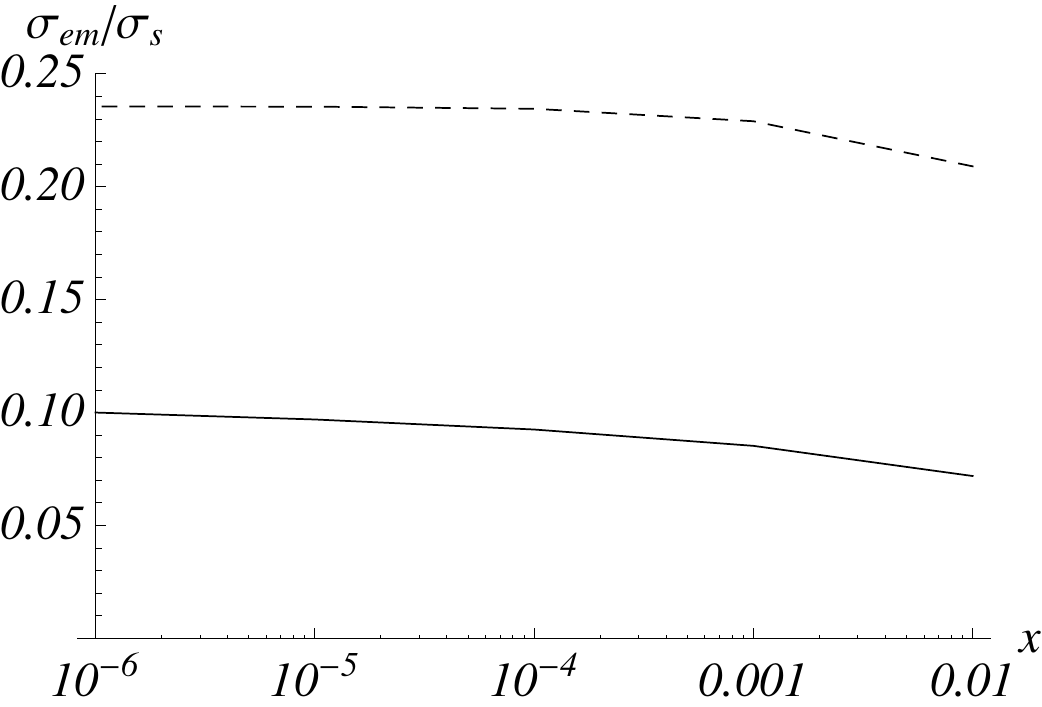} 
 \caption{Ratio of QED and  QCD contributions to the total (solid line) and diffractive (dashed) $\gamma^*A$ cross section as a function of $x$ for gold nucleus at  $Q^2=1$~GeV$^2$.}
\label{fig3}
\end{figure}

\begin{figure}[ht]
     \includegraphics[width=8cm]{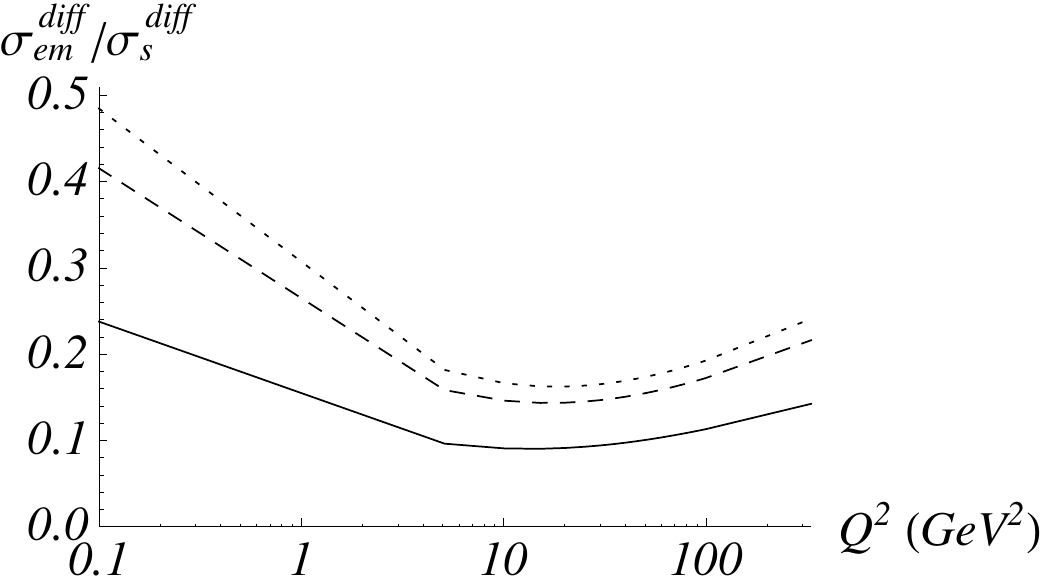}
 \caption{Ratio of QED and  QCD contributions to the diffractive $\gamma^*A$ cross section at $x=10^{-4}$ as a function of $Q^2$ for silver (solid line), gold (dashed line) and uranium (dotted line).  }
\label{fig4}
\end{figure}

 The results are shown in Figs.~\ref{fig1}--\ref{fig4}. All qualitative features agree with our analysis in the previous sections. We can see in \fig{fig1} and \fig{fig4} that at low $Q^2$ the QED correction for uranium nucleus at $x=10^{-4}$ can be as large as 20\% in the total cross section and over 40\% in the diffractive one. It is remarkable that the Coulomb correction is non-negligible even at high $Q^2$. In diffractive cross section, shown in \fig{fig4}, its relative size even increases with $Q^2$, which can be traced back to the extra $\log Q^2$  in \eq{d31}, see \eq{c17}. One should, however, take the results of our calculation at high $Q^2$ with a grain of salt as  the model we are using does not properly account for the DGLAP evolution. A more accurate estimate at high $Q^2$ can be obtained with the model of Ref.~\cite{Bartels:2002cj}. As expected, the relative size of Coulomb corrections increases with the nuclear weight and weakly depends on $x$. 


\section{Summary}

Results presented in this work indicate that Coulomb corrections play an important role in the low $x$ DIS off heavy nuclei in a very wide range of $Q^2$ and $x$. More refined estimates should use  realistic nuclear profiles and sophisticated low $x$ evolution models.  However, they will not change our main conclusion that in order to reliably extract information about the cold nuclear matter structure form the proposed electron-ion collision experiments, one should have the Coulomb correction under  control.

\acknowledgments
This work  was supported in part by the U.S. Department of Energy under Grant No.\ DE-FG02-87ER40371.


\end{document}